\documentclass[a4paper]{article}
\pdfoutput=1

\usepackage[utf8]{inputenc}
\usepackage[T1]{fontenc}
\usepackage[english]{babel}
\usepackage{multirow}
\usepackage[normalem]{ulem}
\usepackage{amsmath}

\usepackage[pdftex,breaklinks=true,bookmarks=true,pdfborder={0 0 0},colorlinks=false]{hyperref}

\usepackage{listings}
\usepackage{amsfonts}
\usepackage{amsmath}
\usepackage{cleveref}
\usepackage{graphicx}
\usepackage[caption=false]{subfig}

\newcommand{\ie}{i.e.~}
\newcommand{\eg}{e.g.~}

\newcommand{\vs}{vs.~}

\usepackage{enumitem}
\setlist{nolistsep}
\usepackage{array}
\usepackage{threeparttable}

\begin{document}

\author{Raphael ‘kena’ Poss\\University of Amsterdam, The Netherlands}
\title{Optimizing for confidence \\ \normalsize Costs and
  opportunities at the frontier between abstraction and reality}

\maketitle

\begin{abstract}
  Is there a relationship between computing costs and the confidence
  people place in the behavior of computing systems? What are the
  tuning knobs one can use to optimize systems for human confidence
  instead of correctness in purely abstract models? This report
  explores these questions by reviewing the mechanisms by which people
  build confidence in the match between the physical world behavior of
  machines and their abstract intuition of this behavior according to
  models or programming language semantics. We highlight in particular
  that a bottom-up approach relies on arbitrary trust in the accuracy
  of I/O devices, and that there exists clear cost trade-offs in the
  use of I/O devices in computing systems. We also show various
  methods which alleviate the need to trust I/O devices arbitrarily
  and instead build confidence incrementally ``from the outside'' by
  considering systems as black box entities. We highlight cases where
  these approaches can reach a given confidence level at a lower cost than
  bottom-up approaches.
\end{abstract}

\setcounter{tocdepth}{2}
\tableofcontents

\clearpage

\section{Introduction}

Many, perhaps most applications of computing systems tolerate some
imperfections in the results of computations. Noise in images, videos
and audio are readily tolerated by human audiences up to a
threshold. The choice of execution order and processing resources
internally to a system can usually be ignored as long as the
externally observable behavior of the system is consistent with
expectations. These examples, with many others, are the symptom of a
\emph{human tolerance for a limited amount of inaccuracy} that can be in
turn exploited by technology providers, because each degree of
flexibility in user expectations can be translated to lower costs,
higher efficiency and better performance.

Meanwhile, a large part of the human interest in computers and the
associated abstract models is the power to define deterministic
algorithms and where proofs of correctness can be formally
ascertained. Declarative semantic systems for functional programming
languages, formal semantic models, formal grammars, formal validation,
automata theory and the like are the tool of theoretical computer
science that equip people with a high degree of trust in their
computing systems, \ie a high level of confidence that the specified
system matches their expectations under some \textsl{a priori}
commonly agreed assumptions.

But this trust is just that: confidence that the \emph{specifications
  and models} of hardware and software components exhibits the desired
properties. But our artefacts never fully match our intuition. More
precisely, our world is one of machines and artefacts that happen to
have observable behaviors, and people around them who attempt to both
prescribe this behavior using programming languages and technology and
describe them using abstract models. At the \emph{boundary} between
machines and models, there exists no rational process that can
ascertain completely in the theoretical domain the behavior of a
machine in the real-world, physical
domain~\cite{russell.12,wittgenstein.22}.

Fortunately, this fundamental limitation does not impede our use of
computers much: we seem content to exploit mutually agreed but
incomplete mechanisms to guide our confidence in the equivalence
between real-world behaviors and models thereof in the abstract
domain. This confidence does not need to be absolute, hence the
tolerance for inaccuracy.

What interests us here is to investigate how this confidence
is built in practice. Although this topic is historically the bread
and butter of professional philosophers, it has tremendous practical
implications in computer science.

For one, the process of building this confidence, via the mechanisms
that ascertain an equivalence between an empirical observation and an
abstract model, has a practical \emph{cost} (time, energy, monetary,
etc). When designing systems to minimize cost, often the costs at the
interface between machines and models are neglected in favor of
complexity costs exclusively in the abstract domain. By understanding
the costs involved at the interface, and taking into account
\emph{how much confidence is actually needed}, new optimization opportunities arise.

The other implication is that
the process of \emph{teaching} people to use, trust and
program computing machines necessarily involves building up trust in
the behavior of man-made systems. It is thus possible to envision that
a better understanding of the confidence build-up process may help
optimize teaching. In an era where the computer architecture industry
is struggling to innovate to break the memory and power
walls~\cite{ronen.01.ieee}, no new understanding that may increase the
competence of newly educated experts should be ignored.

In the present report, we survey the two main approaches currently used to create
this confidence. In \cref{sec:vbc} we review how value-based
equivalence is used to build confidence in models and system
specifications that use a digital encoding of real-world phenomena at
I/O interfaces. These approaches posit axiomatically the trustability
of the I/O interfaces and derive confidence in computing systems
inductively from this basic trust. The contrasting approaches,
reviewed in \cref{sec:cbc}, do not posit that I/O interfaces are to be
trusted, and instead build confidence in the entire behavior of the
computing system ``from the outside'', as it were. In both cases, we
outline the physical costs involved. Meanwhile, we acknowledge that
the issue of cost at the computing interface is not a new field of
study. It has been previously mentioned both for its theoretical and
some practical uses. We review this related work in
\cref{sec:related}. A summary and discussion concludes in \cref{sec:conc}.

\section{Value-based equivalence}\label{sec:vbc}

The most common way to pin a computational model onto phenomena of the
physical world is to rely on input and output devices. The underlying
principles are as follows:
\begin{itemize}
\item for input, the physical phenomenon can be observed and each act
  of observation digitizes the phenomenon into an abstract value.
  This relies on trust that sufficiently similar
  phenomena will cause the device to produce the same digitization.
\item for output, the physical phenomenon can be influenced, and each
  act of influence, also called actuation,  translate an abstract
  value into a physical effect. This relies on trust
  that multiple actuations using the same abstract value will cause
  sufficiently similar physical effects.
\end{itemize}

Using input and output devices, a \emph{value-based equivalence} is
axiomatically posited between abstract values and physical phenomena
(either via observation or actuation, depending on the direction of
the action). The general property of value-based equivalence is that
it establishes equivalence between the physical world and abstract
values independently of what is done with the values: the actions of
input and output are defined independently of the more general
abstract system built on top of them. Confidence in the accuracy of a
model or specification for a system using I/O emerges from both
formal confidence in the correctness of the abstract model in
isolation, and the subjective trust placed in the I/O devices.

\subsection{Value-based equivalence \vs input inaccuracy}

Trust in value-based equivalence using the raw value digitized by an
input device seems weak, since even the most finely manufactured input devices may
report slightly different abstract values for the same (or
sufficiently similar) phenomena.

A common approach  to strengthen this trust is to use an equivalence
threshold: if two digitizations of the same phenomenon $x$ may produce
different abstract values $val_1(x)$ and $val_2(x)$, but $val_1(x)
\ominus val_2(x) < \epsilon$ for some \textsl{a priori} agreed
distance function $\ominus$ and threshold $\epsilon$, then either
$val_1(x)$ or $val_2(x)$ can be used as \emph{representative} of the
phenomenon for the purpose of computing. In practice, sensors
use this principle as follows: a phenomenon is digitized multiple
times in quick succession and the values compared. If the threshold is
not exceeded, then the digitization is accepted and one of the
observations is produced as value. Otherwise, the digitization
is refused and no value is produced.

Trust in the device's accuracy increases when $\epsilon$ can be
decreased without reducing the rate at which observation
actions produce values, \ie its apparent throughput.

Meanwhile, the \emph{physical cost} of performing one input is the
cost of actually performing the raw digitizations \emph{before}
threshold checking, plus the cost of threshold checking itself. For a
given input device, the cost of input rises when reducing $\epsilon$
at constant apparent throughput, because more raw digitizations must
be performed per unit of time. The costs also rises when increasing
throughput at constant $\epsilon$, for the same reason. Conversely,
the cost can be decreased by both increasing $\epsilon$ at constant
apparent throughput (reducing trust) or reducing throughput
requirements at constant $\epsilon$ (reducing performance). This
aspect is especially relevant for energy costs, which have a dynamic
component quadratic with the frequency of raw digitizations and a
static component linear with the size of the value domain.

\subsection{Domain translations and efficiency}

The value domain for representations assumed by algorithm designers
and software integrators is rarely the same as the one envisioned by
the implementers and providers of I/O devices. For example, a camera
sensor may be able to digitize at a resolution of 2000dpi whereas the
algorithm plugged to that sensor only exploits 200dpi worth of
information. On the output side, a servomotor may support angular
controls in increments of .35 degrees (10 bits of resolution) but
exploited by an algorithm that only sets it in one of four positions
(2 bits).

Whenever the abstract model or program specification uses a value
domain different from the domain natively supported by an I/O device,
two questions immediately arise. The first is whether an abstract
model or specification \emph{can} be connected to some specific I/O
devices. The second, assuming they can be connected, is how much the
domain translation costs.

In practice, whether a program assuming some I/O \emph{interface} can
actually run with specific I/O \emph{hardware} is decided on a
case-by-case basis at each deployment, by constructing conversion
functions as necessary. If a conversion function cannot be constructed
using the human resources immediately available, the deployment
fails. To reduce the risk of deployment failure, the design of
modeling and specification (programming) languages is often
self-censored to only define abstract I/O interfaces that are known to
have compatible implementations in hardware. This self-censoring by
language designers is the instrument that creates confidence that the
connection is possible. With this approach, languages are incremented
with new facilities over time as new I/O technology is discovered or
invented. It is an open question whether a universal abstract I/O interface can
be defined that supports any past \emph{and future} I/O technology and
guarantees that a translation always exists between the value domain
at the I/O device and the value domains in the models/programs.

The reason why this discussion is not strictly a computational issue
is that part of the translation may happen in hardware, out of reach
from the abstract domain. Moreover, the reason why the topic of
translation cost matters is that if a model or program uses less
information than provided by an input device, or produces more
information than necessary to actuate an output device, it is possible
to either simplify the input device or simplify the program, incurring
less costs.  When and how this type of optimization is possible
greatly depends on the mismatch between the value domains at the I/O
device and the value domains used by models/programs.

The most trivial case is numeric scaling: the conversion is a
constant-time, constant-space, constant-energy linear scaling between the
value domains. This situation is well-understood and not further
discussed here. What interests us is translations that modify the
\emph{structure} of the value domains. Some examples:

\begin{itemize}
\item \emph{Temporal or spatial ordering} are common domain structures
that are translated by high-level algorithms. For example, an
algorithm may be written to group movies together that share a similar
amount of idle moments (where the image is still and characters don't
speak). The algorithm discards the temporal order of scenes in the
input and considers only the total amount of idle time. Another
algorithm might group images together that share a similar amount of
darkness.  This algorithm discards the spatial order of pixels and
considers only the total amount of darkness.

\item On both input and output, translations over multi-dimensional
  domains commonly \emph{alter their shape and dimensions}.  For
  example, the shape can be discarded entirely: the bits of a
  multi-dimensional value are serialized into a uni-dimensional space,
  such as happens in photographic sensors (the two-dimensional CCD
  grid produces a linear sequence of bits).
\item Partial \emph{projections} are also possible, \ie to a lower
  number of
  dimensions: a program may discard part of three-dimensional
  spatial data to only consider the two-dimensional shadow of objects.

\item Next to altering order and shape, translations may perform
  \emph{integration or derivation} of values. For example, a device
  able to observe geometrical objects in three-dimensional space may
  be used with an integrator to report their volume only
  (integration), or whether their shape is convex (sign of minimum of
  first derivative).
\end{itemize}

Cost-wise, translations have two components: one inherent to the
translation function and one dependent on the particular translated
values. Translations with data-independent costs are usually preferred
as they make cost more easily predictable. For example, the cost of
numeric scalings, translations that only consider ordering, and shape
substitutions over multi-dimensional domains is data-independent. This
is not to mean that translations with data-dependent costs are avoided
entirely: interfaces that perform more complex transformations, even
at unpredictable energy cost, may be preferable to performing the same
translation completely algorithmically because their time cost is
lower.

In any case, as highlighted above, the costs of translation are added
to the cost of digitization when the abstract value domain differs
from the implementation value domain. When the opportunity exists,
costs can be lowered overall by aligning the domains and reducing the
need for translation: either by simplifying devices or the
amount of information manipulated in models and algorithms.

\subsection{Reconfigurable interfaces}

When using fixed I/O devices, the complexity of the mapping between
the device's value domain and the one manipulated by model or program
can only be reduced by changing the model or program.  With
reconfigurable I/O devices, the translation cost can be reduced by
tuning the interface instead.

To illustrate this, consider again the example of the photographic
sensor. The sensor is a grid of photosensitive elements. The total
number of physical elements determines the maximum resolution of the
sensor. If a program requires input images at a lower resolution but operates
the full sensor, all elements are activated and only then some
digitized values are discarded. It would be possible
instead to deactivate some elements in the grid instead, so that the
digitization directly samples the image at the lower
resolution. This uses less energy at the sensors and less time/energy
to down-sample. In other words, by re-configuring the sensor hardware to a
lower resolution, the
desired target value domain is directly reached at a lower cost.

This ability exists in most devices in use today. One can typically
configure the sampling/actuation frequency in signal adapters (video,
audio, network, etc.). Devices also offer tuning knobs for voltage,
resolution, shape, position, threshold. Interestingly, although these
features exist in the interface hardware, it is rarely exposed in
programming interfaces or modeling primitives. We can thus suggest to
study this opportunity as yet another candidate avenue for cost optimization.

\section{Black-box behavior and confidence}\label{sec:cbc}

The previous section has outlined the view where a
model or specification in the abstract domain is connected to real
world, physical behavior via the I/O interfaces of the computing
system. In this approach, human confidence that the behavior of a computing system
interacting with the physical world matches its model or specification
emerges from three factors:
\begin{itemize}
\item trust in the correctness of the abstract model or program taken
  into isolation. This trust can be established in the
  theoretical domain and can become arbitrarily high;
\item for programs, trust in the translation mechanism from the
  abstract specification into bits arrangements in the machine. Again,
  this trust can be constructed mostly in the theoretical domain;
\item trust in the equivalence established by I/O devices between
  physical phenomena outside of the machine and the value domains
  manipulated in the programs/models. This trust is
  established subjectively for each I/O device and system independently.
\end{itemize}

In practice, this approach does not suffice to establish confidence in the
following situations:
\begin{itemize}
\item when a model is constructed \textsl{post hoc} after a system is
  built, without knowledge of how the system is built. How can one
  gain confidence that the model is accurate, \ie that it accurately
  describes and predicts the machine's behavior?
\item when a system is defined using program code that cannot be
  verified, either because the language cannot be analyzed outside of
  the implementation or because the translation mechanisms are not
  trustable. How can one gain confidence that the machine behaves
  according to expectations, \ie that its behavior is sufficiently
  similar to the one specified?
\end{itemize}

In these situations, the machine or apparatus can be studied as a
``black box'' component and confidence in its behavior built by
observations from the outside. In the following sub-sections, we
outline various approaches used to establish \emph{correctness
  judgments} over black-box components. The process of gaining
confidence occurs in the physical world and thus incur costs,
which we also outline.

\subsection{Correctness by fiat}

Fiat correctness occurs when a person (or group) bypasses rational
processes entirely and establishes a correctness judgment and
associated confidence by asserting correctness axiomatically. For
example, numerous voting machines have been stated by fiat by
their manufacturers to be compliant with the model mandated by
regulations.

The immediate physical cost of fiat correctness is the cost of
registering the outcome of the judgment (when required, \eg for
subsequent redistribution). If the judgment is never further used,
the registration costs are obviously nihil. To compensate the
arbitrariness of fiat correctness, its costs are usually extended with
any long-term compensation costs incurred by judgments that
eventually prove incorrect.

\subsection{Judgments modulo physical interaction}

In the natural sciences, correctness judgments and equivalence
relations based on commonalities during physical interactions are
often used. For example, to determine whether two organisms (physical
observations) belong to the same species (model) it is sufficient to
show they can reproduce and yield a fertile offspring (another
physical observation).

The common characteristic of interaction-based equivalence and
correctness judgments is the absence of a direct validation process
between individual objects and the abstract model, combined with the
existence of a validation process between observations of object
interactions and an interaction model.

The costs associated with a single judgment is the cost to identify
real-world objects that \emph{can} interact to yield a judgment,
added to the cost to bring these objects together and cause them to
\emph{actually} interact, added to the cost of observing the resulting
system and deriving the judgment.

Although this empirical process is not a computation overall, it has a
physical cost (time, space, energy, etc.). In particular, it can be
compared to the cost of defining a theoretical model of the
interaction and running a simulation of the model to obtain the
equivalence judgment. For the equivalence to a species model used
above as example, this is nowadays possible using gene sequencing on
the individuals and bio-genetic models to predict reproductive
compatibility. When the organism has a complex genome but sufficiently
fast metabolism, then using an interaction-based judgment may be
cheaper than using the pure computational method. If the genome is
relatively simple and the metabolism slow, the comparison goes the
other way in favor of the computational approach.

Another illustrious example is the use of Amazon's Mechanical Turk to
label product photographs for use in online catalogs. The process of
labeling a photograph is really an operation to ascertain an
equivalence judgment between the image, on one side, and the product
categories recognizable by potential customers on the other side. Were
it done purely computationally, this task would be very expensive:
first a theoretical model must be constructed of the customer's
product categories (a meta-model), then a classifier must be built
sufficiently detailed to distinguish the product models from each
other. In practice, due to economic imbalance between human
populations it is much cheaper to use interaction-based judgments:
bring workers of the Mechanical Turk to interact with the image and
produce an observation reflecting their perception of which model an
image is equivalent to, then use this judgment as a model of what the
typical customer would perceive.

The question of deciding between interaction-based and value-based
judgments for complex real-world phenomena so as to minimize the
cost of judgments is an open problem of computational
science.

\subsection{Correctness by consensus}

Another mechanism used to increase confidence in the
correctness/accuracy of a program/model is to repeat the verification
process. The repetition can occur either over time (over the same
system) or over space (over different systems). The judging entities
can be either other computing systems or humans, using any of the
methods described above. The confidence in the judgment then grows
\emph{not with the number of repetitions, but as the deviation between
  different judgments decreases}, \ie as ``consensus'' is reached.

Within computing systems, correctness by consensus is routinely used
as a mechanism for fault tolerance and recovery; is then called
``redundancy''. However, consensus-based judgments also occur in
other circumstances. Perhaps the most illustrious example is the use
of repeated experiments to validate a theory (model) in the natural
sciences.  This is the foundational process to validate incomplete
models in the modern scientific method.

Another use of correctness by consensus is to increase confidence in
fiat judgments. For example, consensus over fiat correctness is
routinely used to evaluate student performance before delivering a
diploma, where the equivalence must be established between an
observation of that the student did and the committee members'
individual standards for ``sufficient'' and ``insufficient'' work. In
this context, the consensus-based construction is assumed to
compensate for the arbitrariness of judgments of individual committee
members.

Obviously, the cost of correctness by consensus grows with the number
of repetitions.

The reason why consensus-based judgments are relevant here is that
combining consensus with other low-cost black-box judgments may yield
the same confidence at lower cost than trying to construct confidence
from the bottom up as in \cref{sec:vbc}.

For example, consider the
problem of determining routes over dynamic networks (where links
change over time). In this context, gossiping protocols where nodes
with a partial view of their neighbors and possibly inaccurate view
of connectivity typically obtain good routes at a lower cost overall
than an algorithm that would repeatedly compute the best routes using
successive complete models of the network, because the cost of
gathering the complete models at a single location before the global
algorithm can execute is high.

\begin{figure}
\centering
\includegraphics[width=.45\textwidth]{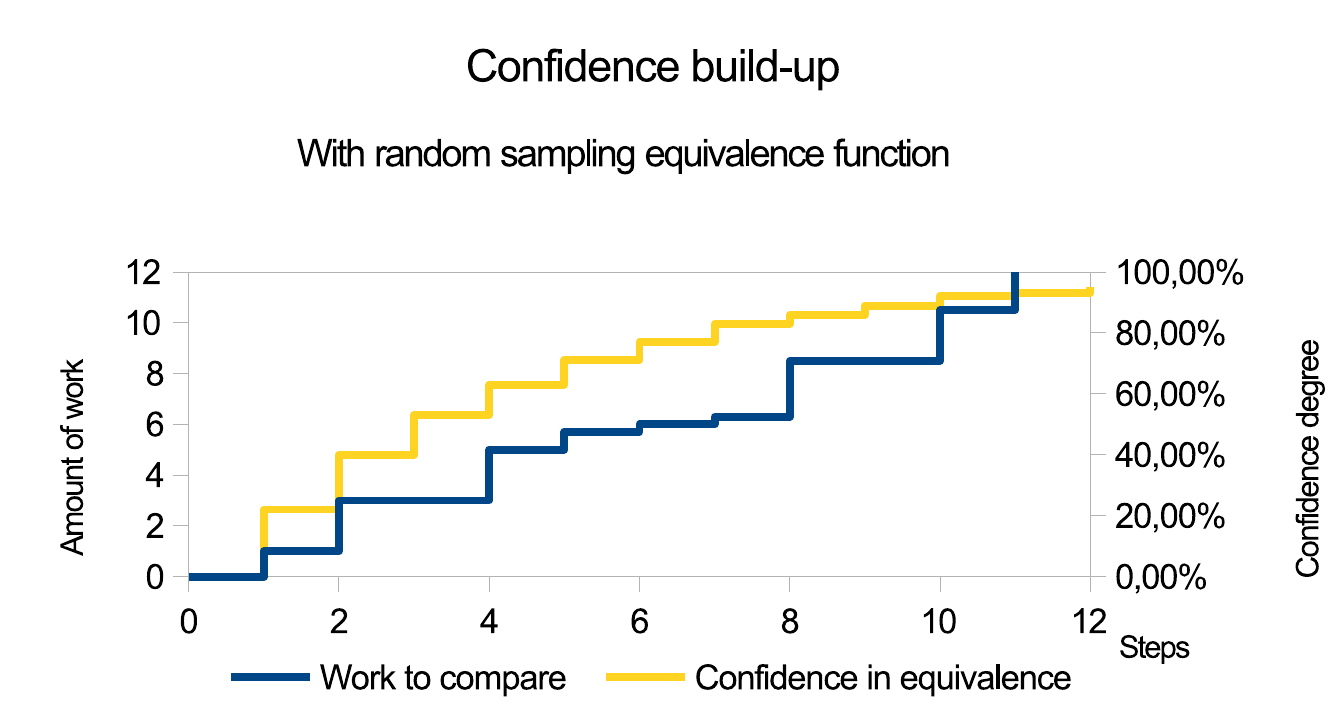} %
\includegraphics[width=.45\textwidth]{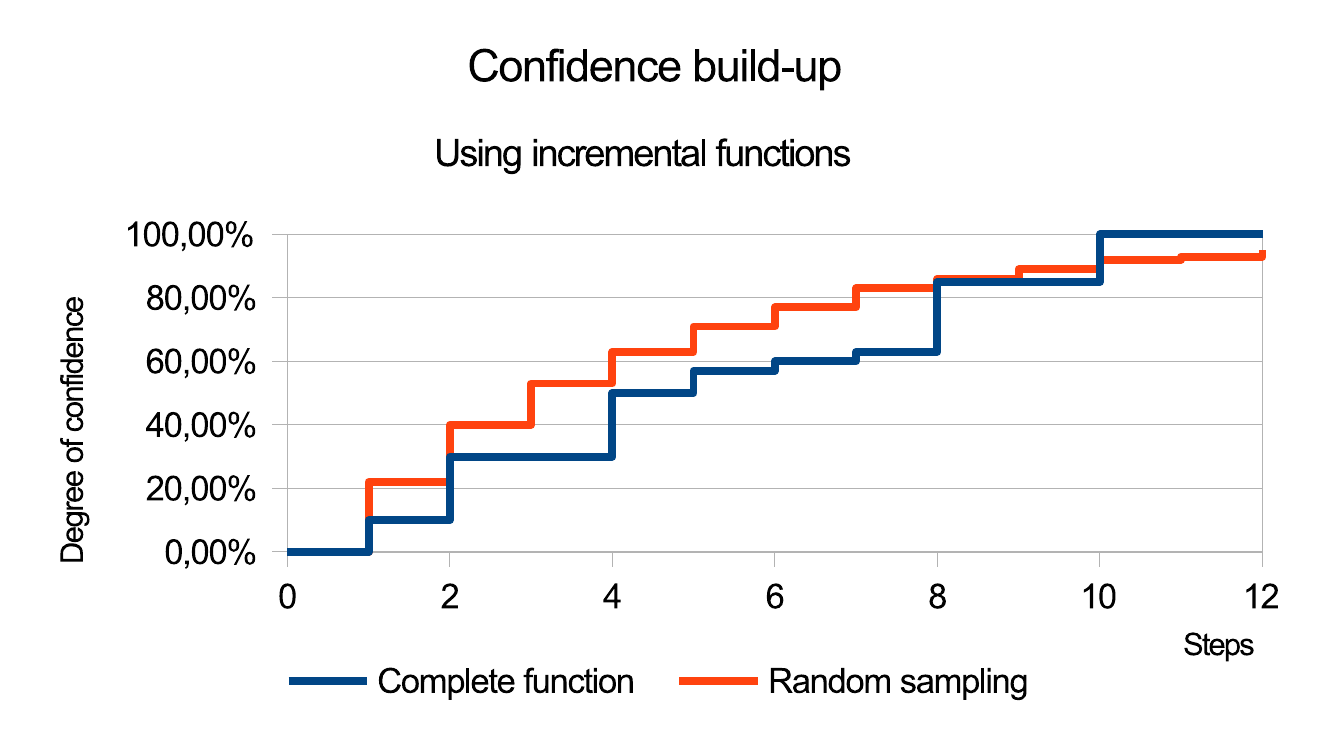}
\caption{Confidence build-up with random sampling and in comparison to
  incremental complete comparisons.}\label{fig:random}
\end{figure}

Another example can be found in situations where a model is compared
to a series of observations, for example to verify its accuracy. If
the total space of potential observations is described in the model
and finite, it is theoretically possible to explore exhaustively this
space by presenting observations one after the other to the comparison
function. In practice however, this space is often large enough that
an exhaustive, in-order comparison is intractable. Instead, one can
pick random samples in the observation space and perform the
comparison for those. As the number of random samples grows,
confidence in the model can grow if the number of dissenting
comparisons stays small. A practical application is the case of
comparisons of digitized images or videos to abstract models. With a
scene digitized as pixels, instead of comparing pixel by pixel over
the entire image size, one can instead pick successive random pixels
and then compare these to the model. As the number of compared pixels
grows, the proportion of the image that has been already compared
increases as well. If they compare equal to their model, confidence in
the equivalence between the whole image and its model grows. Since the
selection is random, some parts may never be compared so the
``maximum'' confidence reachable by a complete comparison cannot be
reached with random sampling. However as shown in \cref{fig:random}
the confidence can reach early on a higher level than a complete
comparison at the equivalent comparison cost.

\section{Related work}\label{sec:related}

Cost trade-offs at the boundary between physical world artefacts and
theoretical models are rarely discussed in scientific circles.  In the
theoretical computer science, one of the more explicit discussions of
costs in machine models is a 1990 survey by P. van Emde
Boas~\cite{emdeboas.90}, where the relationship between algorithms and
space and time costs is discussed for different machine models. In
this survey, the author is careful to mention that cost models for the
behavior of machines must factor the cost of translating real-world
observations to abstract values, \ie the cost of I/O, although these
costs are not discussed further in the survey.

On a more practical side, trade-offs between cost and confidence have
been identified before under the notion of ``approximate
computing''. Research in this area observes that when accounting for
human expectations on the quality of results, one can do away with
complete and accurate solutions and use partial, approximate solutions
instead at a lower cost. The instruments of this approach are, on the
one side, I/O devices with higher error thresholds and on the other
side, algorithms and machine specifications with a limited amount of
non-determinism in the data
path~\cite{sampson.11,esmaeilzadeh.12}. These approaches create
confidence in the overall behavior of approximate systems by bounding
the probability of error in otherwise deterministic
specifications. However they do not discuss or exploit directly the
trade-offs at the boundary between the specification and the actual
physical implementations.

\section{Summary and conclusion}\label{sec:conc}

We have identified various mechanisms by which people build confidence
in their understanding of the behavior of computing systems. We
have distinguished bottom-up approaches which start by trusting
the equivalence between abstract values and real-world phenomena
created by I/O devices. We have highlighted how confidence can be
built from the outside of systems considered as black boxes, for
cases where the bottom-up approach is not practical. While doing so,
we have identified several opportunities for cost optimization:
\begin{itemize}
\item when the value domain of I/O devices does not match the value
  domain in algorithms or models, by either changing the specification
  in the abstract domain or reconfiguring the I/O devices;
\item by combining consensus-based correctness judgments with other
  approaches.
\end{itemize}

Our proposed take-away is the observation that confidence build-ups
using bottom-up approaches are in some cases more expensive than
grouping multiple individual, partial judgments using consensus, when
the goal is to obtain a given target confidence level. This
observation confirms commonplace but often unstated knowledge from
various sub-fields of computer science, in particular those working
with distributed and parallel systems and approximate computing.

\section*{Acknowledgments}
\addcontentsline{toc}{section}{Acknowledgments}

This document reports on thoughts nurtured during
discussions with Alex Shafarenko, Merijn Verstraaten, Sebastian
Altmeyer and Peter van Emde Boas.

\newcommand{\etalchar}[1]{#1} 
\addcontentsline{toc}{section}{References}
\bibliographystyle{is-plainurl}
\bibliography{doc}

\end{document}